\documentclass[12pt, a4paper, twoside]{article}
\usepackage{amsmath}
\usepackage{amsfonts}
\usepackage{amssymb}
\usepackage{amsfonts}

\usepackage{mathtools}
\newenvironment{changemargin}[2]{%
\begin{list}{}{%
\setlength{\leftmargin}{#1}%
\setlength{\rightmargin}{#2}%
}%
\item[]}
{\end{list}}
\usepackage{graphicx}
\usepackage{color}
\usepackage[
top    = 2.75cm,
bottom = 2.50cm,
left   = 1.50cm,
right  = 1.50cm]{geometry}
\usepackage{epstopdf}
\usepackage{pdflscape}
\usepackage{amsmath}
\usepackage{amsmath}
\usepackage{amsmath}
\usepackage{kantlipsum}
\allowdisplaybreaks
\usepackage{setspace}

\usepackage{array}

\newcolumntype{M}[1]{>{\raggedright\arraybackslash}m{#1}}
\newcolumntype{N}{@{}m{0pt}@{}}

\begin{document}
\baselineskip=0.25in
{\bf \LARGE
\begin{changemargin}{-0.5cm}{-0.5cm}
\begin{center}
{Joint remote state preparation (JRSP) of two-qubit equatorial state in quantum noisy channels}\footnote{ {\bf To appear in Phys. Lett. A 2017.}}
\end{center}\end{changemargin}}
\vspace{4mm}
\begin{center}
\large{\bf Adenike Grace Adepoju$^{a,}$}\footnote{\scriptsize E-mail:~ aagrace06@gmail.com}\large{\bf ,} \large{\bf Babatunde James Falaye$^{b,c,\dag,}$}\footnote{\scriptsize E-mail:~ fbjames11@physicist.net; babatunde.falaye@fulafia.edu.ng\\ ${}^\dag$Corresponding Author Tel. No.: 0445531398395}\large{\bf ,} {\large{\bf Guo-Hua Sun$^{d}$}}\footnote{\scriptsize E-mail:~ sunghdb@yahoo.com}\large{\bf ,} {\large{\bf Oscar Camacho-Nieto$^{a,}$}}\footnote{\scriptsize E-mail:~ ocamacho@ipn.mx} \large{\bf and} {\large{\bf Shi-Hai Dong$^{a,}$}}\footnote{\scriptsize E-mail:~ dongsh2@yahoo.com }
\end{center}
{\footnotesize
\begin{center}
{\it $^\textbf{a}$CIDETEC, Instituto Polit\'{e}cnico Nacional, UPALM, CDMX 07700, M\'{e}xico.}\\
{\it $^\textbf{b}$Departamento de F\'isica, Escuela Superior de F\'isica y Matem\'aticas, Instituto Polit\'ecnico Nacional, Edificio 9, UPALM, CDMX 07738, M\'{e}xico.}\\
{\it $^\textbf{c}$Applied Theoretical Physics Division, Department of Physics, Federal University Lafia,  P. M. B. 146, Lafia, Nigeria.}\\
{\it $^\textbf{d}$C\'{a}tedr\'{a}tica CONACyT, CIC, Instituto Polit\'{e}cnico Nacional, UPALM, CDMX 07700, M\'{e}xico.}
\end{center}}
\begin{abstract}
\noindent
This letter reports the influence of noisy channels on JRSP of two-qubit equatorial state. We present a scheme for JRSP of two-qubit equatorial state. We employ two tripartite Greenberger-Horne-Zeilinger (GHZ) entangled states as the quantum channel linking the parties. We find the success probability to be $1/4$. However, this probability can be ameliorated to $3/4$ if the state preparers assist by transmitting individual partial information through classical channel to the receiver non-contemporaneously. Afterward, we investigate the effects of five quantum noises: the  bit-flip noise, bit-phase flip noise, amplitude-damping noise, phase-damping noise and depolarizing noise on the JRSP process.  We obtain the analytical derivation of the fidelities corresponding to each quantum noisy channel, which is a measure of information loss as the qubits are being distributed in these quantum channels. We find that the system loses some of its properties as a consequence of unwanted interactions with environment. For instance, within the domain $0<\lambda<0.65$, the information lost via transmission of qubits in amplitude channel is most minimal, while for $0.65<\lambda\leq1$, the information lost in phase flip channel becomes the most minimal. Also, for any given $\lambda$, the information transmitted through depolarizing channel has the least chance of success.
\noindent
\end{abstract}


Phase damping noise; Depolarizing noise.

{\bf PACs No.}: 03.65.Ud, 03.67.Hk, 03.67.Ac, 03.67.Mn.

\section{Introduction}
\label{sec1}
Suppose we create a pair of particle with a total spin of zero at laboratory $A$ and then take one of the particles to a distant located laboratory $B$. If this particle in laboratory $B$ is found to have clockwise spin on a particular axis, it would be fascinating to see that the spin of the other particle in laboratory $A$, measured on the same axis, would be in counterclockwise direction. This signifies that the quantum state of these particles cannot be elucidated independently. In other words, we refer to the particles as being entangled. Experimental realization of quantum entanglement has been presented in Ref. \cite{T1}. 

The incessant avidity in studying entanglement is as consequence of its outstanding role in several aspects of communication processes. For instance, information stored in quantum system can be teleported from one location to the other with the aid of quantum entanglement which has been previously shared among the sending and receiving locations \cite{T2}. Quantum entanglement is also being utilized in quantum secure direct communication \cite{T3}, quantum dense coding \cite{T4}, hierarchical quantum communication \cite{T5}, etc. In fact, quantum entanglement can be regarded as the heart of quantum communication. With the trend of the search for its applications, one could say that some of its potential relevances are yet to be manifested.

Remote state preparation (RSP) \cite{T7} represents another quantum communication process that utilizes quantum entanglement. RSP is a teleportation of {\it known state}. This is because the sender does not own the particle but all the classical information about the state to be prepared for the remotely located receiver. The sender, say Alice, performs a projective measurement on her qubits in the shared entangled state with the receiver, say Bob and then communicates the result to Bob via classical channel. Depending on the outcome of the measurement, Bob can apply an appropriate quantum gate to reconstruct the original state that Alice intends to transfer from the shared entangled state. This idea was expounded in papers of Lo, Pati and Bennett et al. \cite{T7}. Therein, the communication cost was found to be lower than that of teleportation protocol \cite{T2}. Due to its fascinating properties, RSP has attracted much attentions both within experimental \cite{T9,T10} and theoretical (\cite{T12,T14} and refs. therein) context.

Moreover, RSP has a shortcoming due to the fact that, only one preparer has the complete information about the state to be prepared. Should the preparer not be faithful, then the protocol becomes insecure. Xia et al. \cite{AR1} proposed JRSP in order to tackle this shortcoming in RSP. In the scheme, two parties share information of the known state and jointly collaborate to prepare a particle state for a distant located receiver. In fact, Zhang and co-researchers \cite{T16} have extended this idea to a case of multi-sender and multi-receiver. In line with this, a lot of outstanding works have been reported by so many erudite scholars.  Few of these works can be found in refs. (\cite{T17,T18,T20,RE1,RE2} and refs. therein). These contributions have been made by considering a close quantum system. However, the interaction of real quantum system with surrounding environment is ineluctable.  This unwanted interaction is known as noise which causes gradual loss in coherent of quantum systems. 

Now, what will be the effects of noisy channels on JRSP? Only few studies such as \cite{RE1,RE2,T23,T24} have attempted to deal with related problems and to our best knowledge, JRSP of equatorial state under the influence of noise has not been studied till now. It is therefore the priority purpose of this study to examine this.

The schematic of our presentation is as follows: In section $2$, we review a scheme for JRSP of two-qubit equatorial state using two maximally entangled tripartite GHZ class as quantum channel linking the three parties. In section $3$, we study JRSP process under the influence of five noisy channels, namely; the bit flip, phase flip, bit-phase flip, amplitude-damping, phase-damping and depolarizing channels. Concluding remarks are given in section $4$.

\section{JRSP of two-qubit equatorial states}
For a system of $n$ qubits, the GHZ state can be written as $\left|GHZ\right\rangle=\left(\left|0\right\rangle^{\otimes n}+\left|1\right\rangle^{\otimes n}\right)/\sqrt{2}$. In this paper, we use tripartite GHZ class \cite{T26}; $\left(\left|000\right\rangle+\left|111\right\rangle\right)/\sqrt{2}$, (which evinces non-trivial multipartite entanglement) as quantum channel linking the three parties. GHZ is a type of quantum entanglement which involves at least three subsystems. It is commonly being referred as maximally entangled for some reasons which includes disobeying Bell inequalities maximally. GHZ is totally separable after loss of one qubit unlike W-class which is still entangled with remaining two-qubits. GHZ of three photons and three Rydberg atoms have been observed experimentally \cite{T27}. Using spontaneous parametric down-conversion, three-photon polarization-entangled W state had also been realized experimentally in ref. \cite{T28}. The motivation behind GHZ experiment is due to the fact that GHZ states manifest strong quantum correlations, such that an elegant test of the nonlocality of quantum mechanics is possible \cite{T29}.

Using a couple of the above GHZ state, i.e. $\left|\Phi\right\rangle_{123456}=\left(\left|000\right\rangle+\left|111\right\rangle\right)/\sqrt{2}\otimes\left(\left|000\right\rangle+\left|111\right\rangle\right)/\sqrt{2}$,  we review a protocol for JRSP of two-qubit equatorial state involving two senders and one receiver. Although this problem had been studied recently in ref. \cite{T17,T18}, but this section reviews this problem in a simpler way with different form of computational vector and results of higher success probability. The next section which represents the priority purpose for this study, presents the effect of the aforementioned noises on this protocol. Now, in order to achieve our aim in this section, let us suppose Alice and Bob are the two senders who are located at spatially separated nodes, wish to help the receiver Chika to prepare two-qubit equatorial state of the form 
\begin{equation}
\left|\Psi\right\rangle=\frac{1}{2}\left(e^{\omega_{00}^+}\left|00\right\rangle+e^{\omega_{11}^+}\left|01\right\rangle+e^{\omega_{22}^+}\left|10\right\rangle+e^{\omega_{33}^+}\left|11\right\rangle\right),
\label{E1}
\end{equation}
where $\omega_{nm}^{\pm} ( =i(\alpha_n\pm\beta_m), n,m\in[0,1,2,3])$ denotes the phase shared by Alice and Bob. The $\omega_{00}^\pm$ will eventually be set as zero. Information about this state is known partially to the senders. Let us suppose $\alpha_n$ is known to Alice and $\beta_n$ $(n=1,2,3)$ is known to Bob. This insinuates that there must be a collaboration before the JRSP can be accomplished. We assume that qubits pairs $(1,4)$, $(2,5)$ and $(3,6)$ belongs to Alice, Bob and Chika respectively. Now Alice and Bob perform projective measurements on their respective qubits pairs $(1,4)$ and $(2,5)$ in order to remotely prepare the state of qubit (\ref{E1}) for Chika. For these measurements to be achieved, Alice chooses a set of mutually orthogonal basic vector $\{\left|\zeta_{14}\right\rangle_i, i=1,2,3,4\}$ which are related to computational basis vectors $\{\left|00\right\rangle, \left|01\right\rangle, \left|10\right\rangle, \left|11\right\rangle\}$ in the following form
\begin{eqnarray}
\begin{pmatrix*}\left|\zeta_{14}\right\rangle_1\\ \left|\zeta_{14}\right\rangle_2\\ \left|\zeta_{14}\right\rangle_3\\ \left|\zeta_{14}\right\rangle_4\end{pmatrix*}=
\frac{1}{2}
\begin{pmatrix*}[r]
e^{-i\alpha_0}& e^{-i\alpha_1}& e^{-i\alpha_2}& e^{-i\alpha_3}\\
e^{-i\alpha_0}&-e^{-i\alpha_1}& e^{-i\alpha_2}&-e^{-i\alpha_3}\\
e^{i\alpha_2}& e^{i\alpha_3}&-e^{i\alpha_0}&-e^{i\alpha_1}\\
e^{i\alpha_2}&-e^{i\alpha_3}&-e^{i\alpha_0}& e^{i\alpha_1}\\
\end{pmatrix*}\begin{pmatrix*}[r]\left|00\right\rangle\\ \left|01\right\rangle\\ \left|10\right\rangle\\ \left|11\right\rangle\end{pmatrix*}.
\label{E2}
\end{eqnarray} 
Bob chooses $\{\left|\varrho_{25}\right\rangle_i, i=1,2,3,4\}$ as his measurements basis. These are related to computational basis vector $\{\left|00\right\rangle, \left|01\right\rangle, \left|10\right\rangle, \left|11\right\rangle\}$ as follows:
\begin{eqnarray}
\left(\begin{matrix}\left|\varrho_{25}\right\rangle_1\\ \left|\varrho_{25}\right\rangle_2\\ \left|\varrho_{25}\right\rangle_3\\ \left|\varrho_{25}\right\rangle_4\end{matrix}\right)=
\frac{1}{2}
\begin{pmatrix*}[r]
e^{-i\beta_0}& e^{-i\beta_1}& e^{-i\beta_2}& e^{-i\beta_3}\\
e^{-i\beta_0}&-e^{-i\beta_1}& e^{-i\beta_2}&-e^{-i\beta_3}\\
e^{i\beta_2}& e^{i\beta_3}&-e^{i\beta_0}&-e^{i\beta_1}\\
e^{i\beta_2}&-e^{i\beta_3}&-e^{i\beta_0}& e^{i\beta_1}\\
\end{pmatrix*}\begin{pmatrix*}[r]\left|00\right\rangle\\ \left|01\right\rangle\\ \left|10\right\rangle\\ \left|11\right\rangle\end{pmatrix*}.
\label{E3}
\end{eqnarray}
With these measurements basis, the quantum channel linking the three parties can be written as
\begin{eqnarray}
\left|\Phi\right\rangle_{142536}&=&\frac{1}{2}\left(\left|000000\right\rangle+\left|010101\right\rangle+\left|101010\right\rangle+\left|111111\right\rangle\right)\nonumber\\
&=&\frac{1}{8}\left|\zeta_{14}\right\rangle_1\bigg[\left|\varrho_{25}\right\rangle_1\left(\ e^{\omega_{00}^+}\left|00\right\rangle+e^{\omega_{11}^+}\left|01\right\rangle+e^{\omega_{22}^+}\left|10\right\rangle+e^{\omega_{33}^+}\left|11\right\rangle\right)_{36}\nonumber\\
&&\ \ \ \ \ \ \ \ \ \ +\left|\varrho_{25}\right\rangle_2\left(e^{\omega_{00}^+}\left|00\right\rangle-e^{\omega_{11}^+}\left|01\right\rangle+e^{\omega_{22}^+}\left|10\right\rangle-e^{\omega_{33}^+}\left|11\right\rangle\right)_{36}
\nonumber\\
&&\ \ \ \ \ \ \ \ \ \ +\left|\varrho_{25}\right\rangle_3\left(e^{\omega_{02}^-}\left|00\right\rangle+e^{\omega_{13}^-}\left|01\right\rangle-e^{\omega_{20}^-}\left|10\right\rangle-e^{\omega_{31}^-}\left|11\right\rangle\right)_{36}
\nonumber\\
&&\ \ \ \ \ \ \ \ \ \ +\left|\varrho_{25}\right\rangle_4\left(e^{\omega_{02}^-}\left|00\right\rangle-e^{\omega_{13}^-}\left|01\right\rangle-e^{\omega_{20}^-}\left|10\right\rangle+e^{\omega_{31}^-}\left|11\right\rangle\right)_{36}\bigg]\nonumber\\
&+&\frac{1}{8}\left|\zeta_{14}\right\rangle_2\bigg[\left|\varrho_{25}\right\rangle_1\left(\ e^{\omega_{00}^+}\left|00\right\rangle-e^{\omega_{11}^+}\left|01\right\rangle+e^{\omega_{22}^+}\left|10\right\rangle-e^{\omega_{33}^+}\left|11\right\rangle\right)_{36}\nonumber\\
&&\ \ \ \ \ \ \ \ \ \ +\left|\varrho_{25}\right\rangle_2\left(e^{\omega_{00}^+}\left|00\right\rangle+e^{\omega_{11}^+}\left|01\right\rangle+e^{\omega_{22}^+}\left|10\right\rangle+e^{\omega_{33}^+}\left|11\right\rangle\right)_{36}
\nonumber\\
&&\ \ \ \ \ \ \ \ \ \ +\left|\varrho_{25}\right\rangle_3\left(e^{\omega_{02}^-}\left|00\right\rangle-e^{\omega_{13}^-}\left|01\right\rangle-e^{\omega_{20}^-}\left|10\right\rangle+e^{\omega_{31}^-}\left|11\right\rangle\right)_{36}
\nonumber\\
&&\ \ \ \ \ \ \ \ \ \ +\left|\varrho_{25}\right\rangle_4\left(e^{\omega_{02}^-}\left|00\right\rangle+e^{\omega_{13}^-}\left|01\right\rangle-e^{\omega_{20}^-}\left|10\right\rangle-e^{\omega_{31}^-}\left|11\right\rangle\right)_{36}\bigg]
\nonumber\\
&+&\frac{1}{8}\left|\zeta_{14}\right\rangle_3\bigg[\left|\varrho_{25}\right\rangle_1\left(\ e^{\gamma_{20}}\left|00\right\rangle+e^{\gamma_{31}}\left|01\right\rangle-e^{\gamma_{02}}\left|10\right\rangle-e^{\gamma_{33}}\left|13\right\rangle\right)_{36}\nonumber\\
&&\ \ \ \ \ \ \ \ \ \ +\left|\varrho_{25}\right\rangle_2\left(e^{\gamma_{20}}\left|00\right\rangle-e^{\gamma_{31}}\left|01\right\rangle-e^{\gamma_{02}}\left|10\right\rangle+e^{\gamma_{13}}\left|11\right\rangle\right)_{36}
\nonumber\\
&&\ \ \ \ \ \ \ \ \ \ +\left|\varrho_{25}\right\rangle_3\left(e^{-\omega_{22}}\left|00\right\rangle+e^{-\omega_{33}}\left|01\right\rangle+e^{-\omega_{00}}\left|10\right\rangle+e^{-\omega_{11}}\left|11\right\rangle\right)_{36}
\nonumber\\
&&\ \ \ \ \ \ \ \ \ \ +\left|\varrho_{25}\right\rangle_4\left(e^{-\omega_{22}}\left|00\right\rangle-e^{-\omega_{33}}\left|01\right\rangle+e^{-\omega_{00}}\left|10\right\rangle-e^{-\omega_{11}}\left|11\right\rangle\right)_{36}\bigg]
\nonumber\\
&+&\frac{1}{8}\left|\zeta_{14}\right\rangle_4\bigg[\left|\varrho_{25}\right\rangle_1\left(\ e^{\gamma_{20}}\left|00\right\rangle-e^{\gamma_{31}}\left|01\right\rangle-e^{\gamma_{02}}\left|10\right\rangle+e^{\gamma_{13}}\left|11\right\rangle\right)_{36}\nonumber\\
&&\ \ \ \ \ \ \ \ \ \ +\left|\varrho_{25}\right\rangle_2\left(e^{\gamma_{20}}\left|00\right\rangle+e^{\gamma_{31}}\left|01\right\rangle-e^{\gamma_{02}}\left|10\right\rangle-e^{\gamma_{13}}\left|11\right\rangle\right)_{36}
\nonumber\\
&&\ \ \ \ \ \ \ \ \ \ +\left|\varrho_{25}\right\rangle_3\left(e^{-\omega_{22}}\left|00\right\rangle-e^{-\omega_{33}}\left|01\right\rangle+e^{-\omega_{00}}\left|10\right\rangle-e^{-\omega_{11}}\left|11\right\rangle\right)_{36}
\nonumber\\
&&\ \ \ \ \ \ \ \ \ \ +\left|\varrho_{25}\right\rangle_4\left(e^{-\omega_{22}}\left|00\right\rangle+e^{-\omega_{33}}\left|01\right\rangle+e^{-\omega_{00}}\left|10\right\rangle+e^{-\omega_{11}}\left|11\right\rangle\right)_{36}\bigg],
\label{E4}
\end{eqnarray}
where $\gamma_{nm}=i(-\alpha_n+\beta_m), n,m\in[0,1,2,3]$. After completing the measurements, Alice and Bob send information about their measurements to the receiver Chika via classical channel. As it can be clearly seen from equation (\ref{E4}), if Alice's measurements are $\{\left|\zeta_{14}\right\rangle_j, j=1,2 \}$ and Bob's measurements are $\{\left|\varrho_{25}\right\rangle_j, j=1,2 \}$, then by performing a unitary transformation on qubit state $(3,6)$, Chika can reconstruct the state of the particle which Alice and Bob intend to prepare for her. For instance, let us consider that Alice's measurement is $\left|\zeta_{14}\right\rangle_2$ and Bob's projective measurement is $\left|\varrho_{25}\right\rangle_1$, then the state of qubits pair (3,6) will collapse to $1/2\left(e^{\omega_{00}^+}\left|00\right\rangle-e^{\omega_{11}^+}\left|01\right\rangle+e^{\omega_{22}^+}\left|10\right\rangle-e^{\omega_{33}^+}\left|11\right\rangle\right)$. Applying unitary operation $I\otimes\sigma_z$, Chika can reconstruct the state Alice and Bob intend to prepare for her.  This implies that only 4 out of 16 states are successful. Regarding the measurements of other 12 states, the JRSP fails. Thus, one can infer that the probability of success is $1/4$. However, we can improve this probability of achieving success by considering some special cases as shown in Table 1. 
\begin{table}[h!]
{\scriptsize
\caption{\footnotesize Case 1 in this table shows Alice's measurement outcome, Bob's measurement outcome and recovery transformation utilized by Chika. Case 2 shows how to ameliorate the success probability from $1/4$ to $1/2$ by assuming that Bob sends partial information to Chika via the classical channel. Case 3 shows how to ameliorate the success probability from $1/2$ to $3/4$ by assuming that Alice sends partial information to Chika via the classical channel.}{\vspace*{10pt}{\footnotesize
\begin{tabular}{p{1cm}p{5cm}p{5cm}p{5cm}}\hline\hline
{}&{}&{}&{}\\[1.0ex]
Case& Alice' measurement outcome        &Bob measurement outcome             &Recovery transformation \\[1.0ex]\hline\\
1   &$\left|\zeta_{14}\right\rangle_1$&$\left|\varrho_{25}\right\rangle_1$ &$I\otimes I$            \\[1.0ex]\hline\\
    &                                   &$\left|\varrho_{25}\right\rangle_2$ &$I\otimes\sigma_z$      \\[1.0ex]\hline\\
    &$\left|\zeta_{14}\right\rangle_2$&$\left|\varrho_{25}\right\rangle_1$ &$I\otimes\sigma_z$      \\[1.0ex]\hline\\
    &                                   &$\left|\varrho_{25}\right\rangle_2$ &$I\otimes I$            \\[1.0ex]\hline\\
2   &$\left|\zeta_{14}\right\rangle_1$&$\left|\varrho_{25}\right\rangle_3$ &$e^{\eta_{02}^+}\left|00\right\rangle\left\langle 00\right|+e^{\eta_{13}^+}\left|01\right\rangle\left\langle01\right|-\newline
																				e^{\eta_{20}^+}\left|10\right\rangle\left\langle 10\right|-e^{\eta_{31}^+}\left|11\right\rangle\left\langle11\right|$            \\[1.0ex]\hline\\
    &                                   &$\left|\varrho_{25}\right\rangle_4$ &$e^{\eta_{02}^+}\left|00\right\rangle\left\langle 00\right|-e^{\eta_{13}^+}\left|01\right\rangle\left\langle01\right|-\newline
																				e^{\eta_{20}^+}\left|10\right\rangle\left\langle 10\right|+e^{\eta_{31}^+}\left|11\right\rangle\left\langle11\right|$            \\[1.0ex]\hline\\
    &$\left|\zeta_{14}\right\rangle_2$&$\left|\varrho_{25}\right\rangle_3$ &$e^{\eta_{02}^+}\left|00\right\rangle\left\langle 00\right|-e^{\eta_{13}^+}\left|01\right\rangle\left\langle01\right|-\newline
																				e^{\eta_{20}^+}\left|10\right\rangle\left\langle 10\right|+e^{\eta_{31}^+}\left|11\right\rangle\left\langle11\right|$            \\[1.0ex]\hline\\
    &                                   &$\left|\varrho_{25}\right\rangle_4$ &$e^{\eta_{02}^+}\left|00\right\rangle\left\langle 00\right|+e^{\eta_{13}^+}\left|01\right\rangle\left\langle01\right|-\newline
																				e^{\eta_{20}^+}\left|10\right\rangle\left\langle 10\right|-e^{\eta_{31}^+}\left|11\right\rangle\left\langle11\right|$            \\[1.0ex]\hline\\
3   &$\left|\zeta_{14}\right\rangle_3$&$\left|\varrho_{25}\right\rangle_1$ &$e^{\zeta_{02}^+}\left|00\right\rangle\left\langle 00\right|+e^{\zeta_{13}^+}\left|01\right\rangle\left\langle01\right|-\newline
																				e^{\zeta_{20}^+}\left|10\right\rangle\left\langle 10\right|-e^{\zeta_{31}^+}\left|11\right\rangle\left\langle11\right|$          \\[1.0ex]\hline\\
    &                                   &$\left|\varrho_{25}\right\rangle_2$ &$e^{\zeta_{02}^+}\left|00\right\rangle\left\langle 00\right|-e^{\zeta_{13}^+}\left|01\right\rangle\left\langle01\right|-\newline
																				e^{\zeta_{20}^+}\left|10\right\rangle\left\langle 10\right|+e^{\zeta_{31}^+}\left|11\right\rangle\left\langle11\right|$          \\[1.0ex]\hline\\                                
    &$\left|\zeta_{14}\right\rangle_4$&$\left|\varrho_{25}\right\rangle_1$ &$e^{\zeta_{02}^+}\left|00\right\rangle\left\langle 00\right|-e^{\zeta_{13}^+}\left|01\right\rangle\left\langle01\right|-\newline
																				e^{\zeta_{20}^+}\left|10\right\rangle\left\langle 10\right|+e^{\zeta_{31}^+}\left|11\right\rangle\left\langle11\right|$            \\[1.0ex]\hline\\
		&																		&$\left|\varrho_{25}\right\rangle_2$ &$e^{\zeta_{02}^+}\left|00\right\rangle\left\langle 00\right|+e^{\zeta_{13}^+}\left|01\right\rangle\left\langle01\right|-\newline
																				e^{\zeta_{20}^+}\left|10\right\rangle\left\langle 10\right|-e^{\zeta_{31}^+}\left|11\right\rangle\left\langle11\right|$            \\[1.0ex]\hline
\end{tabular}
\begin{itemize}
\item $\eta_{nm}^\pm=i(\beta_n\pm \beta_m)$
\item $\zeta_{nm}^\pm=i(\alpha_n\pm\alpha_m)$
\end{itemize}
}}}
\end{table}
\section{JRSP of two-qubit equatorial state in noisy environment}
The review in the last section has been carried out within the framework of closed quantum system. However, a realistic quantum system will ineluctably interact with the environment. Without the result of the influence of noise factor on this protocol, this study is not yet complete. This motivates the study in this section. Quantum systems lose their properties as a ramification of undesirable interactions with the environment. Experimental and theoretical works on effects of quantum noises on RSP and JRSP have been reported in refs. (\cite{T30,T31,T32,T33} and refs). To our best knowledge, there has been no report so far on JRSP of two-qubit equatorial state in noisy environment. It is therefore the priority purpose of this section to scrutinize this. We shall consider the aforementioned quantum noisy channels as models for the noise. The analytical expression for fidelity in each cases will be derived in order to determine information loss as quantum information is been transmitted through these channels. 

\subsection{JRSP in Bit flip, phase flip and bit-phase flip channels}
\begin{figure*}[!h]
\centering \includegraphics[height=140mm, width=165mm]{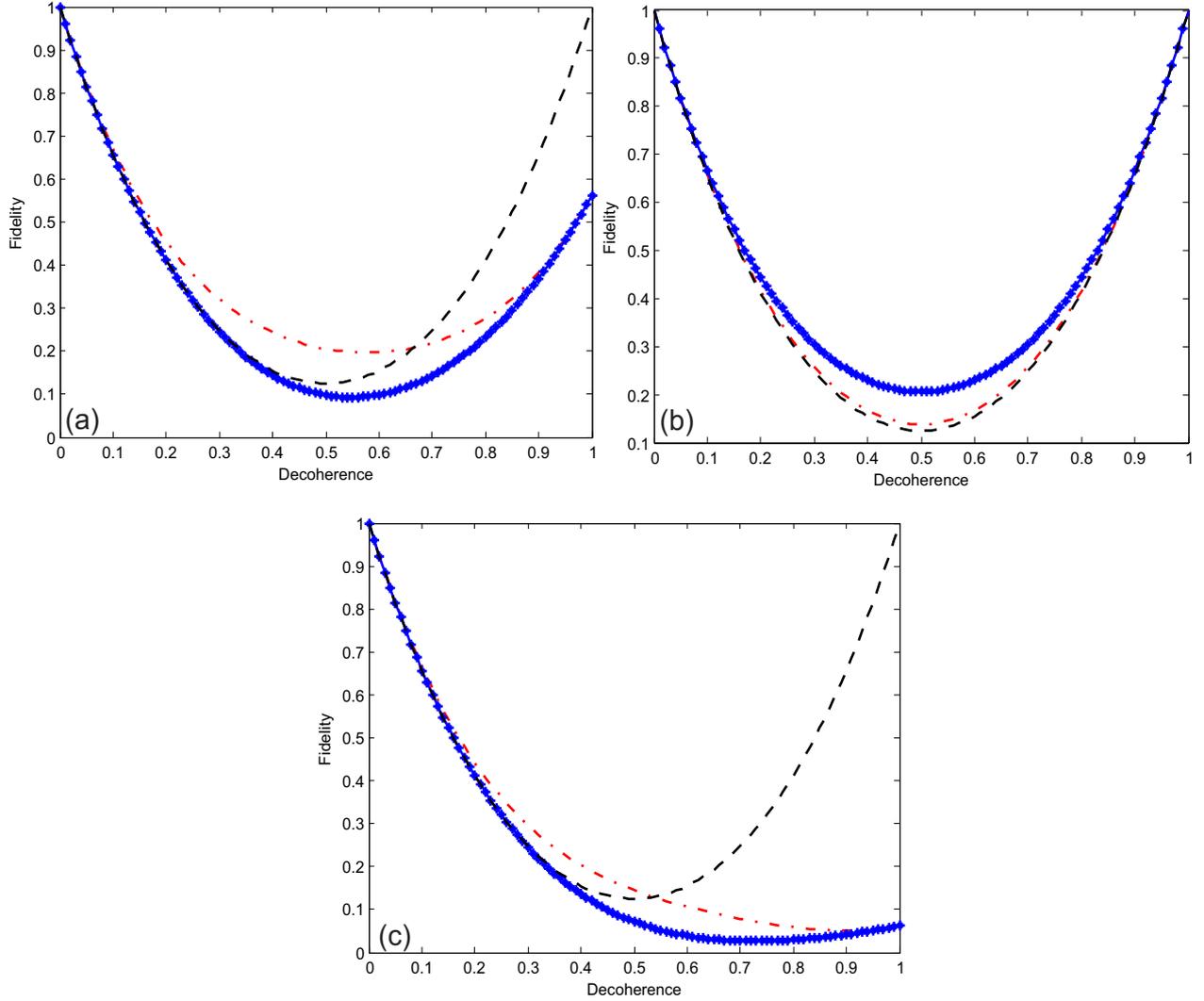}
\caption{\protect\small Plots of fidelities as a function of decoherence rate of bit flip, phase flip channel, and bit-phase flip channels. In (a), we take $\alpha_t=\beta_t=30$. In (b), we choose $\alpha_t=\beta_t=180$. In (c), we set $\alpha_t=\beta_t=300$. The line marker ``$-\cdot$" represents bit flip while ``- -" represents the phase flip and ``-*" represents bit-phase flip. The line markers are in respect to MATLAB notations. In all cases, $t=1, 2, 3$.}
\label{fig1}
\end{figure*}

The Pauli $X$ and $Z$ matrices are also called as bit flip (b-f) and phase flip (p-f) matrices. The Pauli $X$ matrix maps $\left|0\right\rangle$ to $\left|1\right\rangle$, and vice-versa, justifying the name {\it bit flip}. The $Z$ matrix leaves $\left|0\right\rangle$ invariant, and maps $\left|1\right\rangle$ to $-\left|1\right\rangle$, substantiating the name {\it phase flip}. The bit-phase flip (bp-f) is the combination of a phase flip and a bit flip. In this subsection, we study the influences of b-f, p-f and bp-f on JRSP of two-qubit equatorial state.  The general behavior of these channels are characterized by the following set of Kraus operators \cite{T34}
\begin{subequations}
\begin{eqnarray}
E_0^{b-f}&=&\sqrt{1-\lambda}\left[\begin{matrix}1&0\\0&1\end{matrix}\right], \ \ \ \ \ \ \ \ \ \ \ \ \ \ \ 
E_1^{b-f}=\sqrt{\lambda}\left[\begin{matrix}0&1\\ 1&0\end{matrix}\right],\\
E_0^{p-f}&=&\sqrt{1-\lambda}\left[\begin{matrix}1&0\\0&1\end{matrix}\right], \ \ \ \ \ \ \ \ \ \ \ \ \ \ \ 
E_1^{p-f}=\sqrt{\lambda}\left[\begin{matrix}1&0\\ 0&-1\end{matrix}\right],\\
E_0^{bp-f}&=&\sqrt{1-\lambda}\left[\begin{matrix}1&0\\0&1\end{matrix}\right], \ \ \ \ \ \ \ \ \ \ \ \ \ \
E_1^{bp-f}=\sqrt{\lambda}\left[\begin{matrix}0&-i\\ i&0\end{matrix}\right],
\label{E5}
\end{eqnarray}
\end{subequations}
where $\lambda\ (0\leq\lambda\leq 1)$ denotes the noise parameter. Now, let us begin with bit flip channel. Since qubit pair $(3,6)$ is not transmitted through noisy channel, ergo, we shall only consider the influence of this noisy channel on qubit pairs $(1,4)$ and $(2,5)$ in the shared entangled state. Thus, we have
\begin{equation}
\mathcal{B}^f(\rho)=\sum_{i,j}E_i^{b-f^1}\otimes E_i^{b-f^4}\otimes E_j^{b-f^2}\otimes E_j^{b-f^5}\rho\left(E_i^{b-f^1}\otimes E_i^{b-f^4}\otimes E_j^{b-f^2}\otimes E_j^{b-f^5}\right)^\dag,
\label{E6}
\end{equation}
where $\rho$ denotes the density matrix of the shared state, i.e., $\rho=\left|\Phi\right\rangle\left\langle\Phi\right|$ with $i,j$ being element of $\{0,1\}$ and the superscripts $(^{1425})$ represent the action of operator $E$ on which qubit.  From section 2, our review reveals that JRSP will only be successful if Alice's measurements are $\left|\zeta_{14}\right\rangle_{1,2}$ and Bob's measurements are $\left|\varrho_{25}\right\rangle_{1,2}$ except if the two senders assist the receiver as we have explained before. This is also appurtenant under noisy condition. It then insinuates that the fidelity in failure cases cannot be considered. Thus, the shared entangled state (in the basis 123456) becomes a mixed state as a ramification of interaction with the environment. The output density matrix can be determined (in the basis $3,6$)  from which we obtain an analytical expression for the fidelity as 
\begin{eqnarray}
F^{b-f}&=&(1-\lambda)^4+\frac{\lambda^4}{16}\Big[e^{\bar{\omega}_{03}}+e^{\bar{\omega}_{12}}+e^{\bar{\omega}_{21}}+e^{\bar{\omega}_{30}}\Big]^2+\frac{\lambda^2(1-\lambda)^2}{16}\Big[e^{\eta_{03}^-}+e^{\eta_{12}^-}+e^{\eta_{21}^-}+e^{\eta_{30}^-}\Big]^2\nonumber\\
&&+\frac{\lambda^2(1-\lambda)^2}{16}\Big[e^{\zeta_{03}^-}+e^{\zeta_{12}^-}+e^{\zeta_{21}^-}+e^{\zeta_{30}^-}\Big]^2,
\label{E10}
\end{eqnarray}
where we have used $\bar{\omega}_{nm}={\omega}_{nn}-{\omega}_{mm}$ for mathematical simplicity. Our analytic expression shows that the fidelity depends on phase shared and noise parameter. For $\lambda=0$, $F=1$, which represents a perfect joint remote state preparation. However, for $\lambda=1$, $F=1/{16}\Big[e^{\bar{\omega}_{03}}+e^{\bar{\omega}_{12}}+e^{\bar{\omega}_{21}}+e^{\bar{\omega}_{30}}\Big]^2$. Repeating the same calculation procedure, we obtain the fidelity of JRSP for phase flip channel as $F^{p-f}=\lambda^4+(1-\lambda)^4$ and the fidelity of JRSP for bit-phase flip channel as
\begin{eqnarray}
F^{bp-f}&=&(1-\lambda)^4+\frac{\lambda^4}{16}\Big[e^{\bar{\omega}_{03}}+e^{\bar{\omega}_{12}}+e^{\bar{\omega}_{21}}+e^{\bar{\omega}_{30}}\Big]^2+\frac{\lambda^2(1-\lambda)^2}{16}\Big[e^{\eta_{03}^-}-e^{\eta_{12}^-}-e^{\eta_{21}^-}+e^{\eta_{30}^-}\Big]^2\nonumber\\
&&+\frac{\lambda^2(1-\lambda)^2}{16}\Big[e^{\zeta_{03}^-}-e^{\zeta_{12}^-}-e^{\zeta_{21}^-}+e^{\zeta_{30}^-}\Big]^2.
\label{E11}
\end{eqnarray}
In this case, for $\lambda=1$, $F^{bp-f}=F^{b-f}$. We examine the variation of the fidelities as a function of decoherence rate of bit flip, phase flip channel, and bit-phase flip channels in figure \ref{fig1}. From figures \ref{fig1}(a-c), it can be seen that $F^{p-f}(\lambda=0)=F^{p-f}(\lambda=1)=1$. This result insinuates that if the noisy channel is phase-flip channel, then the JRSP scheme would be successful provided that the decoherence is at its peak. In figure \ref{fig1}(a), we take $\alpha_t=\beta_t= 30$. It can be seen that the fidelities of the channels dwindle with increasing decoherence until $\lambda=0.5$ which could be deemed as a critical point for phase damping channel where the variation pattern changes from inverse proportionality to direct proportionality and consequently becoming a well.

The turning point can be observed for other noisy channels by taking $\alpha_t=\beta_t=180$ as in figure \ref{fig1}b. The vertex of each curve falls on the axis of the symmetry which then yields $\lambda=0.5$ as the turning point. Moreover, the scenario is quite different in figure 1c by considering $\alpha_t=\beta_t=300$. Generally speaking, the analysis of figures \ref{fig1}(a-c) evinces that the fidelity of the quantum system can be permuted via variation of phase information.
\subsection{JRSP in amplitude-damping channel}
Amplitude-damping noise represents one of the valuable decoherence noise which provides us with description of energy-dissipation effects due to loss of energy from quantum state. For instance, within the framework of weak Born-Markov approximation, the model for amplitude-damping noise is very resourceful in elucidating spontaneous emission of a photon by a two-level system into an environment of photon at a lilliputian temperature. Experimentally, this noisy channel can be realized via a Sagnac-type interferometer with an additional beam splitter to facilitate the tracing out of environment qubit. The general behavior of this noise is characterized by the following set of Kraus operators \cite{T34,T35}
\begin{equation}
E_0^A=\left[\begin{matrix}1&0\\0&\sqrt{1-\lambda_{A}}\end{matrix}\right], \ \ \ \mbox{and}\ \ \ \ E_1^A=\left[\begin{matrix}0&\sqrt{\lambda_{A}}\\0&0\end{matrix}\right],
\label{E12}
\end{equation}
where $\lambda_{A}(0\leq\lambda_{A}\leq 1)$ represents the decoherence rate which characterizes the probability error of amplitude-damping when a particle passes through a noisy channel. Since qubit pair $(3,6)$ is not transmitted through noisy channel, thus the effect of the amplitude-damping noise on the shared entangled state can be represented as follows:
\begin{equation}
\mathcal{A}(\rho)=\sum_{i,j}E_i^{A^1}\otimes E_i^{A^4}\otimes E_j^{A^2}\otimes E_j^{A^5}\rho\left(E_i^{A^1}\otimes E_i^{A^4}\otimes E_j^{A^2}\otimes E_j^{A^5}\right)^\dag,
\label{E13}
\end{equation}
where $i,j\in\{0,1\}$ and the superscripts $(^{1425})$ denote the action of operator $E$ on which qubit. The shared state which becomes a mixed state after particles distribution. The density matrix of the output state (in the basis $36$) can be found and in order to determine closeness of the final state to the initial state, we employ the fidelity . Thus, we have
\begin{eqnarray}
F=\frac{1}{16}\bigg[1+2(1-\lambda_{A})+(1-\lambda_{A})^2\bigg]^2+\frac{\lambda_{A}^4}{16}e^{2\bar{\omega}_{30}}+\frac{\lambda_{A}^2(1-\lambda_{A})^2}{16}\left[e^{2\zeta_{30}}+e^{2\eta_{30}}\right],
\label{E17}
\end{eqnarray}
which is a function of phase parameter and the decoherence rate. For $\lambda_{A}=0$, then $F=1$ which denotes a perfect JRSP while for $\lambda_{A}=1$, $F=1/16(1+e^{2\bar{\omega}_{30}})$. Figure 2 (a) shows the plot of fidelity given by Eq. (\ref{E17}) as a function of decoherence rate. We observe that the fidelity dwindles as $\lambda_A$ increases. Furthermore, as a subplot ($\alpha$), we have the contour plot of the fidelity as a function of $\lambda_A$ and $\alpha_1$ ($rad$). This figure reveals that varying the phase factor has no effect on the fidelity.
\subsection{JRSP in phase-damping channel}
In this subsection, we examine the effect of phase-damping channel on the JRSP of two-qubit equatorial state. This channel provides a simple model for decoherence. It also describes the loss of quantum information without loss of energy which denotes one of the important feature of this channel. In fact, one could say that this channel models a noise effect that is  exclusively quantum mechanical in nature. A typical example of this is randomly scattering of photon as it transverses through a waveguide. The energy eigenstate does not vary as a function of time, instead it amasses phase which commensurates with eigenvalue. Consequently, the limited information regarding the relative phase between energy eigenstates is lost when the evolution time is not known. The behavior of this noise is characterized by the following set of Kraus operators \cite{T33,T34,T35,T36}
\begin{equation}
E_0^P=\sqrt{1-\lambda_{P}}\left[\begin{matrix}1&0\\0&0\end{matrix}\right],\ \ \ \ E_1^P=\sqrt{\lambda_{P}}\left[\begin{matrix}1&0\\0&0\end{matrix}\right]\ \ \ \mbox{and}\ \ \ \ E_2^P=\sqrt{\lambda_{P}}\left[\begin{matrix}0&0\\0&1\end{matrix}\right],
\label{E18}
\end{equation}
where $\lambda_{P}(0\leq\lambda_{P}\leq 1)$ denotes the decoherence rate for the phase-damping noise. Again, we consider the fact that qubit pair $(3,6)$ is not transmitted through noisy channel, we express the effect of the phase-damping noise on the shared entangled state as
\begin{equation}
\mathcal{P}(\rho)=\sum_{i,j}E_i^{P^1}\otimes E_i^{P^4}\otimes E_j^{P^2}\otimes E_j^{P^5}\rho\left(E_i^{P^1}\otimes E_i^{P^4}\otimes E_j^{P^2}\otimes E_j^{P^5}\right)^\dag.
\label{E19}
\end{equation}
The closeness of the final state to the initial state can be calculated as
\begin{eqnarray}
F=(1-\lambda_{P})^4+\frac{\left(1-\lambda_{P}\right)^2\lambda_{P}^2}{4}+\frac{\lambda_{P}^4}{8},
\label{E22}
\end{eqnarray}
which is independent of phase parameters but depends on the decoherence rate. For $\lambda_{P}=0$, $F=1$ which symbolizes a perfect JRSP. However, for $\eta_{_P}=1$, $F=1/8$.
\begin{figure*}[!h]
\centering \includegraphics[height=70mm, width=165mm]{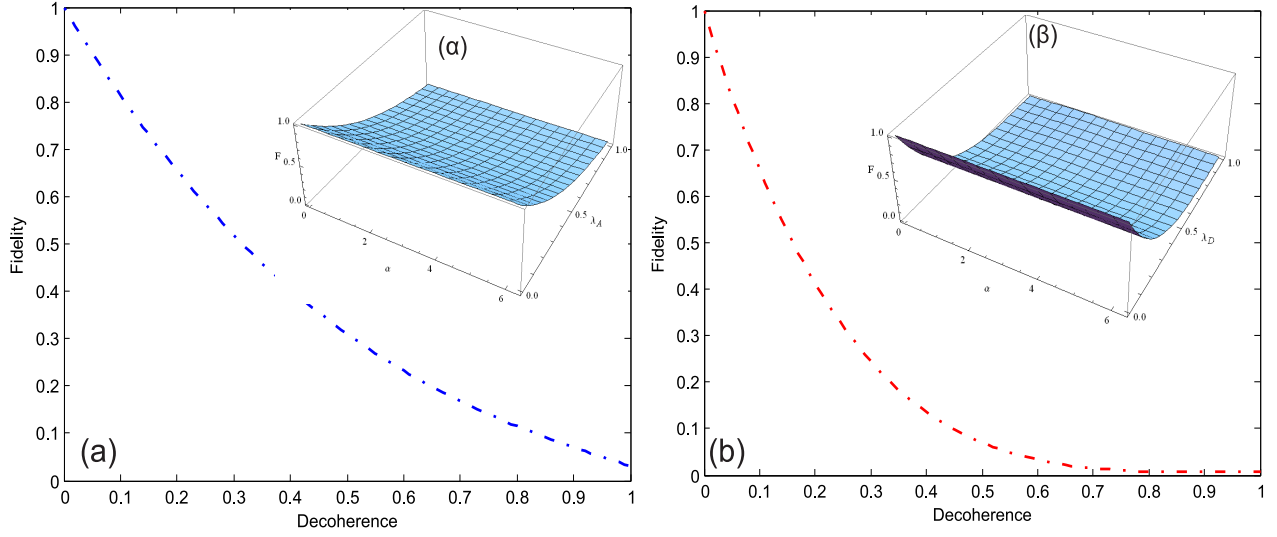}
\caption{\protect\small In (a), we show the plot of fidelity as a function of decoherence rate of amplitude-damping channel. We take $\alpha_t=\beta_t=300$. The subplot ($\alpha$) shows the three-dimensional surface plot of fidelity as a function of decoherence rate and phase information. (b) Same as (a) but for depolarizing channel.}
\label{fig2}
\end{figure*}
\subsection{JRSP in depolarizing channel}
In this subsection, we examine the effect of depolarizing noise on the JRSP of two-qubit equatorial state. Depolarizing channel can be described as a model that has outstanding symmetry properties which introduce white noise. The behavior of this noise is characterized by the following set of Kraus operators \cite{T33,T34,T35,T36}
\begin{equation}
E_0^D=\sqrt{1-\lambda_{D}}{\bf1},\ \ \ \ E_1^D=\sqrt{\frac{\lambda_{D}}{3}}{\bf\sigma_1},\ \ \ \ E_2^D=\sqrt{\frac{\lambda_{D}}{3}}{\bf\sigma_2}\ \ \ \mbox{and}\ \ \ \ E_3^D=\sqrt{\frac{\lambda_{D}}{3}}{\bf\sigma_3},
\label{E23}
\end{equation}
where $\lambda_{D}(0\leq\lambda_{D}\leq 1)$ denotes the decoherence rate for the depolarizing noise. Here, we also consider the fact that only qubit pair $(1,4)$ and $(2,5)$ are transmitted through noisy channel. Consequently, the effect of the depolarizing noise on the shared entangled state can be expressed as
\begin{equation}
\mathcal{D}(\rho)=\sum_{i,j}E_i^{D^1}\otimes E_i^{D^4}\otimes E_j^{D^2}\otimes E_j^{D^5}\rho\left(E_i^{D^1}\otimes E_i^{D^4}\otimes E_j^{D^2}\otimes E_j^{D^5}\right)^\dag,
\label{E24}
\end{equation}
where $i,j\in\{0,1,2,3\}$. Following the same calculation procedure of subsections $3.1-3.3$, we obtain the fidelity as
\begin{eqnarray}
F&=&(1-\lambda_D)^4+\lambda_D^2\left(\frac{(1-\lambda_D)^2}{72}-\frac{\lambda_D^2}{648}\right)\left[\left(e^{\eta_{12}^-}+e^{\eta_{21}^-}\right)^2+\left(e^{\zeta_{12}^-}+e^{\zeta_{21}^-}\right)^2\right]+\frac{\lambda_D^4}{81}\nonumber\\
&&+\frac{\lambda_D^4}{648}\left[\left(e^{\bar{\omega}_{12}}+e^{\bar{\omega}_{21}}\right)^2+\left(e^{\bar{\omega}_{03}}+e^{\bar{\omega}_{12}}+e^{\bar{\omega}_{21}}+e^{\bar{\omega}_{30}}\right)^2\right],
\label{E25}
\end{eqnarray}
which is a function of phase parameter and the decoherence rate. For $\lambda_D=0$, then $F=1$ which is perfect JRSP. In figure 2(b) we examine the variation of fidelity for depolarizing channel as a function of $\lambda_D$. This figure shows that as the depolarizing channel becomes more decoherence, the fidelity reduces. Also, from the subplot ($\beta$), we can deduce that altering the phase factor has no effect on the fidelity.

\begin{figure*}[!h]
\centering \includegraphics[height=70mm, width=165mm]{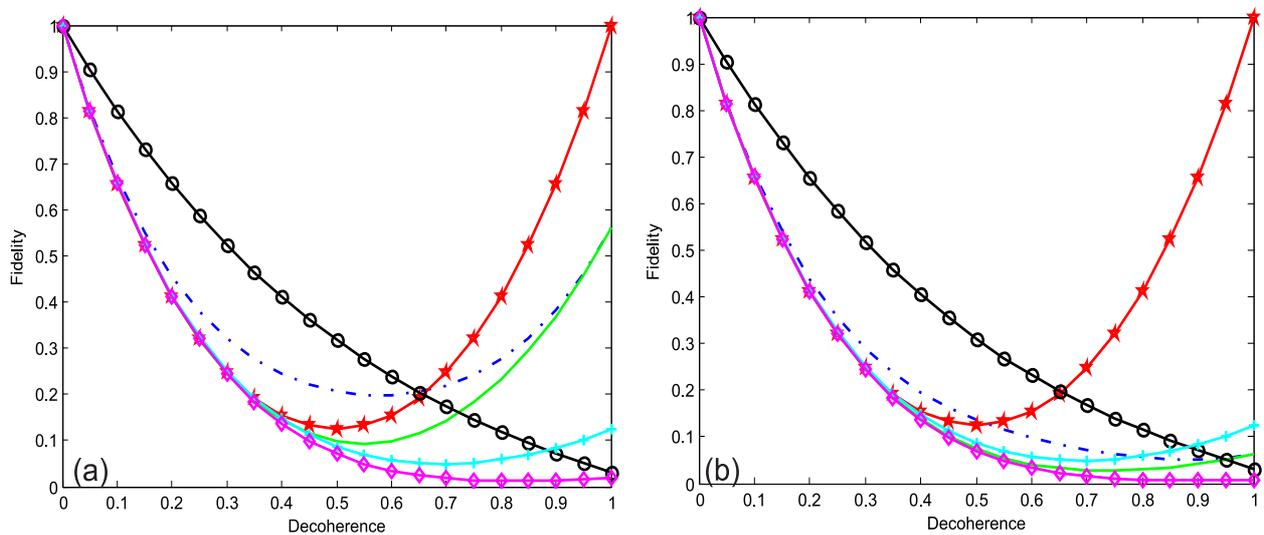}
\caption{\protect\small Plots of fidelities as a function of decoherence rate of bit flip, phase flip, bit-phase flip, amplitude-damping, phase-damping and depolarizing channels. In (a), we take $\alpha_t=\beta_t=30$. In (b), we take $\alpha_t=\beta_t=300$. The line marker ``$-\cdot$" , ``-" , ``-pentagram", ``-o", ``-+", ``-d"  represent bit flip, phase flip, bit-phase flip, amplitude-damping, phase-damping and depolarizing channels respectively. In all cases, $t=1, 2, 3$.}
\label{fig3}
\end{figure*}

Figure \ref{fig3} shows the variation of fidelities as a function of decoherence rate for various noisy channels. This figure shows that amplitude damping channel is the most decoherence. This can be seen by considering a particular fidelity in figure \ref{fig3}(a), say $F=0.6$, the corresponding value of $\lambda_A\approx0.25$ where as for other noisy channel it is about $0.15$. Also, we can infer from this figure that for $0<\lambda<0.65$, the information lost via transmission of qubits in amplitude channel is most minimal. Transition occur at $\lambda=0.65$ and for $0.65<\lambda\leq1$, the information lost in phase flip channel becomes the most minimal while for any given $\lambda$, the information transmitted through depolarizing channel has the least chance of success. Moreover, figure \ref{fig3}(b) shows that only bit flip and bit-phase flip channels are sensitive to variation in phase factor.

\section{Conclusions}
Noise is a noteworthy impediment to the development of practical quantum information processing devices. Conceptual comprehension and control of such noise processes would expedite the constructions of many useful quantum information processing systems. In this letter we have reported the influence of noisy channels on joint remote state preparation of two-qubit equatorial state. In order to realize this, first, we examine a scheme for joint remote state preparation of two-qubit equatorial state. We employ two tripartite Greenberger-Horne-Zeilinger (GHZ) entangled states as the quantum channel linking the parties. We found the probability of success to be $1/4$. However, this probability has been ameliorated to $3/4$ by assuming that the state preparers assist by transmitting individual partial information through classical channel to the receiver non-contemporaneously. Afterward, we investigated the effects of five quantum noises: the  bit-flip noise, bit-phase flip noise, amplitude-damping noise, phase-damping noise and depolarizing noise on the JRSP process.  We obtain the analytical derivation of the fidelities corresponding to each quantum noisy channel, which is a measure of information loss as the qubits are being distributed in these quantum channels. We found that the system loses some of its properties as a consequence of unwanted interactions with environment. Our numerical computations reveal that for $0<\lambda<0.65$, the information lost via transmission of qubits in amplitude channel is most minimal while for $0.65<\lambda\leq1$, the information lost in phase flip channel becomes the most minimal. Also, for any given $\lambda$, the information transmitted through depolarizing channel has the least chance of success. Moreover, if the noisy channel is phase-flip channel, then the JRSP scheme would be successful provided that the decoherence is at its peak. The result obtained in this study will be useful for making improvements to real implementation of quantum secure communication. This study is another boosterish evidence that justifies quantum entanglement as key resource in quantum information science and it also represents the continuation of our recent studies \cite{T33,T37}. We hope that the current study will inspire furtherance in the future by considering multi-preparer with a multi-receiver. 

\section*{Acknowledgments}
We thank the referees for the positive enlightening comments and suggestions, which have greatly helped us in making improvements to this paper. This work was partially supported by 20160978-SIP-IPN, Mexico.

\end{document}